\documentclass[twocolumn, superscriptaddress,nofootinbib,floatfix, amsfonts, aps]{revtex4}

\usepackage{setspace,ulem, epsfig,amssymb,amsfonts,amsmath,mathtools,bm,color,xcolor,graphicx,braket,adjustbox,esint,upgreek}

\begin{document}
\begin{spacing}{1.0}
\title{Open quantum system approach for heavy quark thermalization}
\author{Zhuoxuan Xie} 
\author{Baoyi Chen}\email{baoyi.chen@tju.edu.cn}
\affiliation{Department of Physics, Tianjin University, Tianjin 300350, China}

\date{\today}

\begin{abstract}
We treat heavy quark as an open quantum system in the hot medium, and rederive 
the Stochastic Schr\"odinger Equation (SSE)  
from the full Schr\"odinger equation for both heavy quarks and the medium. 
We apply the SSE to the dynamical evolutions of heavy quarks (as a system) 
in the static hot medium (as an environment).  
Heavy quarks interact with the medium via random scatterings, which 
exchange the momentum and phase factor randomly between 
two wave functions of the system and the environment. The exchange of momentum and 
phase factor results in 
the transition between different eigenstates of the system. These are 
included via an external stochastic potential in the Hamiltonian of SSE. 
Stochastic wave functions of heavy quarks are 
evolved with the stochastic external potential.  
The mean wave functions and the 
corresponding momentum distributions of heavy quarks 
are obtained after the ensemble average over a large 
set of stochastic wave functions. 
We present the thermalization of heavy quarks in the static medium with different 
coupling strength.

\end{abstract}
\maketitle

\section{Introduction}

A deconfined matter consisting of quarks and gluons called 
``Quark-Gluon Plasma'' (QGP) is believed to be produced in 
the relativistic heavy-ion collisions~\cite{Bazavov:2011nk}. 
Heavy quarks are produced in  
initial parton hard scatterings in the nuclear collisions.  
The thermal production in the medium is negligible 
due to the large thresh-hold of heavy quark mass.   
Therefore, heavy quarks and quarkonium have been proposed to be clean probes 
for the early stage of the hot deconfined 
medium in heavy-ion collisions~\cite{Matsui:1986dk,Thews:2000rj,Greco:2003vf,
Grandchamp:2003uw,Yan:2006ve,Liu:2010ej,Blaizot:2017ypk,Chen:2019qzx,Yao:2018sgn,Yao:2020xzw}.
Heavy quarks suffer significant energy in the medium 
~\cite{vanHees:2007me,Zhao:2017yan,Rapp:2018qla} via both 
 elastic collision~\cite{Qin:2010mn} and the medium-induced parton 
radiation~\cite{Baier:1996kr, Qin:2007rn}. 
Assume small momentum transfer in each 
collision, heavy quark evolution can be treated as a 
Brownian motion. The energy loss 
of heavy quarks have been studied with classical models such as 
the Langevin equation~\cite{He:2011qa,Cao:2013ita} and the 
transport equation~\cite{Cao:2016gvr}.  
 
Heavy quarks subjected to the quantum environment can also be described 
with an open quantum system approach~\cite{ref:qsm1,Brambilla:2020qwo,Blaizot:2017ypk}, 
such as the Stochastic Schr\"odinger Equation model which treats medium interactions 
as stochastic potentials in the Hamiltonian~\cite{Katz:2015qja,Akamatsu:2018xim}. 
In analogy with Brownian particles diffusing in a classical 
phase space along classical stochastic trajectories, 
quantum states of the system (heavy quark) 
diffuse in a Hilbert space with stochastic modifications 
of the wave function. 
In classical models, 
the interaction between heavy quarks and the medium is encoded 
in the diffusion coefficient of the 
Langevin equation~\cite{He:2012df,Chen:2017duy}. 
While in quantum models, 
the interaction potential can be included directly 
in the Hamiltonian of heavy quarks. 
 
In this work, we treat heavy quarks as an open quantum system and the QGP as the 
quantum environment. We start with the 
full Schr\"odinger equation including two wave functions of 
the system and the environment. 
Tracing out the degree of freedom of the environment, 
we obtain a reduced Schr\"odinger equation with an external time-dependent 
stochastic potential describing the random interactions with the medium. 
In the Markovian case where 
the time-correlation in the stochastic potential become a delta function, SSE has been 
proved to be equivalent to the Lindblad equation after the ensemble average over 
a set of stochastically-evolved wave functions~\cite{ref:sse1}. 
The computing cost in the framework of evolving wave functions directly 
is much smaller 
than the case of Lindblad equation evolving the density matrix. 
In this work,  
we obtain SSE and apply to the thermalization of heavy quarks in the static medium. 

The work is organized as follows. In Section II, we derive the 
SSE for heavy quarks in the Markovian limit. 
In Section III, we present the evolutions of heavy quark momentum distribution in the 
static medium. 
A summary and discussion about further applications of SSE 
in heavy-ion collisions are given in Section IV.

\section{Open Quantum System Approach}
To treat heavy quarks as an open quantum system in the 
medium, we 
follow the Ref.\cite{ref:sse1} and write the full Schr\"odinger equation for both 
system and the environment (with natural units $\hbar=c=1$), 

\begin{align}
i {d\ket{\Phi(t)}\over dt} = (\hat{H}_{S} + \hat{H}_{B} + \lambda \hat{W}(t))\ket{\Phi(t)}
\end{align}
where $\hat{H}_S$ and $\hat{H}_B$ 
is the Hamiltonian of the system and the thermal bath (environment), 
respectively. $\hat{W}(t)$ is the interaction between the system 
and the environment. $\lambda$ is 
the coupling constant. $\ket{\Phi(t)}$ is the full wave function of the system 
and the environment. In the interaction picture, the Schrodinger equation becomes, 
\begin{align}
\label{eq-schinter}
i {d\ket{\Phi_I(t)}\over dt} =  \lambda \hat{W}_I(t)\ket{\Phi_I(t)}
\end{align}
where the full wave function in the interaction picture 
consists of two parts: the wave 
functions of the system and the environment, 
$\ket{\Phi_I(t)}=\sum_n \ket{\varphi_I^{n}(t)}
\otimes \ket{n}$. The projection operators 
$\hat{P}={1}_S\otimes \ket{l}\bra{l}$ 
and $\hat{Q}=1_S\otimes\sum_{n\neq l} \ket{n}\bra{n}$ are introduced, which 
can separate one specific state of the environment $\ket{l}$ from other states 
$\sum_{n\neq l}\ket{n}$~\cite{ref:project1,ref:project2,ref:sse1}. 
Two operators satisfy the relation $\hat{P}+\hat{Q}={1}$. 
Apply $\hat{P}$ 
and $\hat{Q}$ projection operators on the Schr\"odinger equation Eq.(\ref{eq-schinter}) 
respectively, one can separate the full Schr\"odinger equation into two equations, 
\begin{align}
\label{eq-sch-p1}
i dP\ket{\Phi_I (t)} 
&= \lambda\hat{P}\hat{W}_I(t)\hat{P}\ket{\Phi_I (t)}dt + \lambda\hat{P}\hat{W}_I(t)\hat{Q}\ket{\Phi_I (t)}dt\\
\label{eq-sch-p2}
i d\hat{Q}\ket{\Phi_I (t)} 
&= \lambda\hat{Q}\hat{W}_I(t)\hat{P}\ket{\Phi_I (t)}dt + \lambda\hat{Q}\hat{W}_I(t)\hat{Q}\ket{\Phi_I (t)}dt
\end{align}
From Eq.(\ref{eq-sch-p2}) one can get the time evolution of the selected 
state $\hat{Q}\ket{\Phi_I(
\Delta t)}$ to be,  
\begin{align}
\label{eq-qstate}
\hat{Q}\ket{\Phi_I (\Delta t)} 
&= (1-i\lambda\hat{Q}\hat{W_I}(0) \hat{Q})\ket{\Phi_I (0)})\Delta t \nonumber \\
&\quad - i\lambda\hat{Q}\hat{W_I}(0)\hat{P}\ket{\Phi_I (0)} \Delta t
\end{align}
where $\Delta t$ is the time step of the evolution. 
Pluge Eq.(\ref{eq-qstate}) back into the Schro\"odinger equation Eq.(\ref{eq-sch-p1}), 
one can get the equation for the projected state $\hat{P}\ket{\Phi_I(t)}$. 
In the weak coupling limit where $\lambda$ is small, 
we neglect the high orders of $\lambda$ to get a simplified equation, 
\begin{align}
\label{eq-pphi1}
i d\hat{P}\ket{\Phi_I (t)} 
&= \lambda \hat{P}\hat{W_I}(t)\hat{P}\ket{\Phi_I (t)}dt \nonumber \\
&\quad + \lambda \hat{P}\hat{W_I}(t)\hat{U}(t)\hat{Q}\ket{\Phi_I (0)} dt 
+\mathcal{O}(\lambda^2)
\end{align}
where the evolution operator satisfies the relation 
$d\hat{U}(t)=-i\lambda \hat{Q}\hat{W}_I(t)\hat{Q}\hat{U}(t)dt$. Introduce the 
communication relation $[\hat{U}(t),\hat{Q}]=0$ and take the expansion of $\hat{U}(t)$ over 
$\lambda$ in 
the weak coupling limit, we get the equation for the selected state  
$\hat{P}\ket{\Phi_I(t)}$, 
\begin{align}
\label{eq-sim-fsch}
i d{P}\ket{\Phi_I (t)} 
=& \lambda {P}{W_I}(t)P\ket{\Phi_I (t)}dt +\lambda P W_I(t) Q\ket{\Phi_I(0)}
\nonumber \\ &\quad+\mathcal{O}(\lambda^2)
\end{align}
which describes the evolutions of the system 
wave function with the environment located in one specific state 
selected by the operator $\hat{P}$. 
Now we take a linear form of the interaction term 
$\hat{W}_I(t)=\sum_\alpha \hat{V}_\alpha \otimes \hat{B}_\alpha$ in the SSE. 
After multiplying $\bra{l}$ in Eq.(\ref{eq-sim-fsch}), 
the first term on the R.H.S becomes zero when we redefine the operator $\hat{B}_\alpha$ 
to get the relation $\bra{l}\hat{B}_\alpha\ket{l}=0$.
The dynamical equation for $l$-th mode wave function $\varphi_I^l(t)$ becomes, 
\begin{align}
\label{eq-int8}
&\bra{l}i d\hat{P}\ket{\Phi_I (t)} = 
id\varphi^l_I(t)\nonumber \\
&= \lambda \sum_{\alpha,n\neq l} \bra{l} \hat{B_\alpha}\ket{n}\hat{V_\alpha}\ket{\varphi_I^n (0)}dt +\mathcal{O}(\lambda^2) \nonumber \\
&=\lambda \sum_{\alpha,n\neq l} \bra{l} \hat{B_\alpha}\ket{n}\hat{V_\alpha}e^{-{\beta\over 2}
(\epsilon_n-\epsilon_l)}e^{i(\theta_n-\theta_l)}\ket{\varphi_I^l (0)}dt +\mathcal{O}(\lambda^2) 
\end{align}
where the $\lambda$-terms describe the influence of different environment 
modes ($n\neq l$) on the wave function $\ket{\varphi_I^{l}(0)}$. 
In this work, we consider a system in the equilibrated medium and the initial wave function 
of the system is taken as a pure state $\phi(0)$. 
The initial full wave function of the system and the environment 
is written as $\ket{\Phi_I(0)}=\sum_n \ket{\phi_I(0)}\otimes 
\sqrt{{e^{-\beta\epsilon_n}\over Z_B}} e^{i\theta_n}\ket{n}$ with $\beta\equiv1/T$ where 
$T$ is the temperature. 
The thermal particles consisting of the medium are assumed to satisfy 
Boltzmann distribution. $\epsilon_n$ 
is the energy of the medium (which are thermal massless particles) 
located at $n$-th eigenstate. 
$Z_B=\rm{Tr_B} e^{-\beta T}$ is the trace. 
$\theta_n$ is the phase factor in the wave function of the environment.  
With these setups, the initial conditions of the 
Eq.(\ref{eq-int8}) becomes $\ket{\varphi_I^n(0)}= 
\ket{\phi_I(0)}
\sqrt{{e^{-\beta\epsilon_n}\over Z_B}} e^{i\theta_n}$. From the 
definition of the wave function $\ket{\varphi_I^{n}(t)}$, 
we can see that it 
includes both the system wave function $\ket{\phi_I(t)}$ and also 
the information of the environment states, 
which enters into the evolution of the system 
wave function. Considering Eq.(\ref{eq-sim-fsch}) describes the evolution of 
a specific wave function $\ket{\varphi_I^l}$, 
we do an average over different medium eigenstates to obtain the evolution 
of the state $\ket{\phi_I(t)}$, which is considered as the wave function of the system. 
In the Schr\"odinger picture, 
the reduced Schr\"odinger equation with the interactions between the system and the 
environment is, 
\begin{align}
\label{eq-schf2}
i{d\ket{\phi(t)}\over dt}=\hat{H}_s\ket{\phi(t)}
+\lambda \sum_\alpha \gamma_\alpha(t) \hat{V}_\alpha\ket{\phi(t)}
\end{align}
where $\gamma_\alpha$ is a stochastic noise term, 
\begin{align}
\gamma_\alpha(t)= \sum_{l,n\neq l} {1\over \sqrt{Z_B}} 
\bra{l} \hat{B}_\alpha\ket{n}e^{-{\beta\over 2}
\epsilon_n}e^{i(\theta_n-\theta_l)}
\end{align}
and $\hat{H}_s$ is the Hamiltonian of the system without interactions. 
$\theta_n$ is the 
phase factor transferred from the medium wave function to the system wave function. 
In the Markovian limit, 
the stochastic 
interaction is treated as a white noise in the Hamiltonian.   
the time correlation of the random phase factors satisfy a delta function, 
\begin{align}
\langle \theta_i(t)\theta_j(t^\prime)\rangle = \Theta\delta_{ij}\delta(t-t^\prime)
\end{align}
where $i$ and $j$ are the index of different medium eigenstates. 
The random phase factor is uniformly distributed in $[-\pi,\pi]$ 
which gives $\Theta={\pi^2\over 3}$. 
The term with random phase makes the system evolve towards 
an uniform 
distribution over all the states. The other damping 
term $e^{-{\beta\over 2}\epsilon_n}$ makes the system evolve 
towards the low energy states~\cite{Akamatsu:2018xim}.
The competition between two factors draw the system towards the 
thermalization distribution of the medium.

Here quark-gluon plasma as an environment is assumed to be an ideal massless gas, 
where thermal particles are located in N kinds of discrete 
eigenstates. The wave function of the medium is expressed as 
$\ket{n_1, n_2,..., n_N}$.  $n_\xi (\xi=1,...,N)$ 
represents the number of thermal 
particles located at the $\xi$-th eigenstate with 
the momentum $\xi\cdot \Delta p_{\rm en}$. $\Delta p_{\rm en}$ is 
the momentum gap between the eigenstates. 
The system wave function can also be expressed with a series of 
discrete eigenstates $\ket{\phi(t)}=\sum_{i=0}^M c_i(t) \ket{i}$, 
with the momentum step between those eigenstates 
to be $\Delta p_{\rm sys}$.  $M$ is the total number of the system 
eigenstates in the numerical calculations. 
In the interaction, heavy quark can obtain the momentum  
 $\xi\cdot \Delta p_{\rm en}$ from the medium by  
absorbing a thermal particle at the state $\ket{n_\xi}$ or 
dump energy to the medium 
by emitting a corresponding particle. 
The medium wave function is changed from $\ket{n_\xi}$ to 
$\ket{n_\xi +1}$ state correspondingly.  
There is also transitions between the $i$-th and $j$-the 
eigenstates in  heavy quark wave function under the rule   
$|j-i|\cdot \Delta p_{\rm sys}= \xi\cdot \Delta p_{\rm en}$. 
We introduce the interaction term between the heavy quark and the medium to be, 
\begin{align}
\label{eq-Winter}
\hat{W}
=\sum_{i=1}^{M-1}\sum_{j=i+1}^{M}  \hat{a}_{\xi}\ket{j}\bra{i} + \hat{a}^\dag_{\xi}\ket{i}\bra{j}
\end{align}
where
the annihilation operator $\hat{a}_{\xi}$ 
and the creation operator $\hat{a}_{\xi}^\dagger$ 
change the medium state $\ket{n_1,n_2,...n_\xi...}$ 
as 
$\hat{a}_{\xi}\ket{n_\xi} =\sqrt{n_\xi}\ket{n_\xi-1}$ 
and $\hat{a}^{\dag}_{\xi}\ket{n_\xi}=\sqrt{n_\xi+1}\ket{n_\xi+1}$. 
Heavy quark wave function is also changed from $i$-th to $j$-th state 
according to the rule 
$\xi\cdot \Delta p_{\rm en}=|j-i|\cdot \Delta p_{\rm sys}$. 
In order to label all the variables in SSE with the index of the system eigenstates, 
we perform the replacement $\epsilon_\xi=\epsilon_{ij}$ and $\ket{n_\xi}=\ket{n_{ij}}$ according 
to the transition rule 
$\xi\Delta p_{\rm en}=|j-i|\Delta p_{\rm sys}$. 
The lowering operator $\hat{a}_{ij}$ reduce the number of thermal particles located at 
the medium eigenstate $\ket{n_{\xi}}$ and heavy quark is shifted from $i$-th to 
$j$-th state ($j>i$).  
The term with the raising operator $\hat{a}^\dag_{ij}$ 
represents the process of heavy quarks dumping energy to the environment by emitting 
a thermal particle.  
The SSE for heavy quark evolution as an open quantum system 
is written as, 

\begin{align}
\label{eq-sse-complete}
&i {d\ket{\phi(t)}\over dt} 
= 
 \hat{H_s}\ket{\phi(t)} \nonumber \\ 
&+\frac{\lambda}{\sqrt{Z_B}} \sum_{i=1}^{M-1}\sum_{j=i+1}^{M}
[\bra{n_{ij}-1}\hat{a}_{ij}\ket{n_{ij}} e^{-\frac{\beta}{2}\epsilon_{n_{ij}}} 
e^{i\theta_{ij}^\prime} c_i\ket{j}\nonumber \\
 &\quad + \bra{n_{ij}+1}\hat{a}_{ij}^\dagger \ket{n_{ij}} e^{-\frac{\beta}{2}
\epsilon_{n_{ij}}} 
e^{i\theta_{ij}^{\prime\prime}} c_j\ket{i}] 
\end{align}
where $\hat{H_s}$ 
is the Hamiltonian of heavy quarks without interactions. 
Heavy quark mass is taken as 1.5 GeV. 
$\theta_{ij}^\prime=\theta_{n_{ij}} - \theta_{n_{ij}-1}$ and 
$\theta_{ij}^{\prime\prime}=\theta_{n_{ij}} - \theta_{n_{ij}+1}$ are the 
random phase factors transfered from the medium wave function to the heavy quark wave 
function in each interaction. 
$i$ on the L.H.S of the equation represents the imaginary number. 
While $i,j$ in the summation 
and subscript are the index of the heavy quark state. 
$\epsilon_{n_{ij}}$ 
is the energy of the $n_{ij}$ thermal particles 
located at the medium $\xi$-th eigenstate where 
$\xi\Delta p_{\rm en}=|j-i|\Delta p_{\rm sys}$. $\lambda$ is the 
coupling constant. 

\section{Numerical Results}

As a preliminary study,  
we study the heavy quark kinetic thermalization in 
the static uniformly-distributed medium. 
The medium state is represented 
with discrete eigenstates with the momentum gap $\Delta p_{\rm en}=0.2$ GeV/c. 
Assume the medium is an ideal gas, 
we can get the number of thermal particles located at each  
eigenstate. We initialize the total number of 
thermal particles at different states 
to be $\sum_i n_i=100$ where $n_i$ satisfies Boltzmann distribution with 
the temperature $T=1$ GeV. 
Heavy quark wave function is written as a quantum superposition of eigenstates, 
where the momentum gap is 
$\Delta p_{\rm sys}=0.12$ GeV/c. The total number of system eigenstates is chosen to 
be $M=30$. The time step of the numerical evolution in Eq.(\ref{eq-sse-complete}) 
is taken to be $dt=0.001$ fm/c. 
We have tested that different steps of the time and momentum do not affect 
the results evidently. 
We consider different cases of the coupling strength between heavy quark and the 
environment. Inspired by the heavy quark potential at finite 
temperature~\cite{Kaczmarek:2002mc, Lafferty:2019jpr}, 
we take the coupling strength to be 
$\lambda=0.3$ and  $0.5$ to perform the preliminary calculations. 
The exact determination of the $\lambda$ 
will be left in the future work.  
At each time step, heavy quark exchange momentum and phase factor with the medium by 
emitting or absorbing a massless thermal particle. 
As an open quantum system, 
the wave function of heavy quarks is not normalized due to the stochastic interactions, 
which is normalized by hand at each time step. 
With a large set of stochastic wave functions, we obtain a mean wave function 
$\overline{\phi(t)}$ after the ensemble average. $|\bra{i}\ket{\overline{\phi(t)}}|^2$ 
can be interpreted as a probability of 
heavy quarks located at the state with the momentum  
$i\cdot \Delta p_{\rm sys}$. 

\begin{figure}[h]
\includegraphics[width=0.49\textwidth]{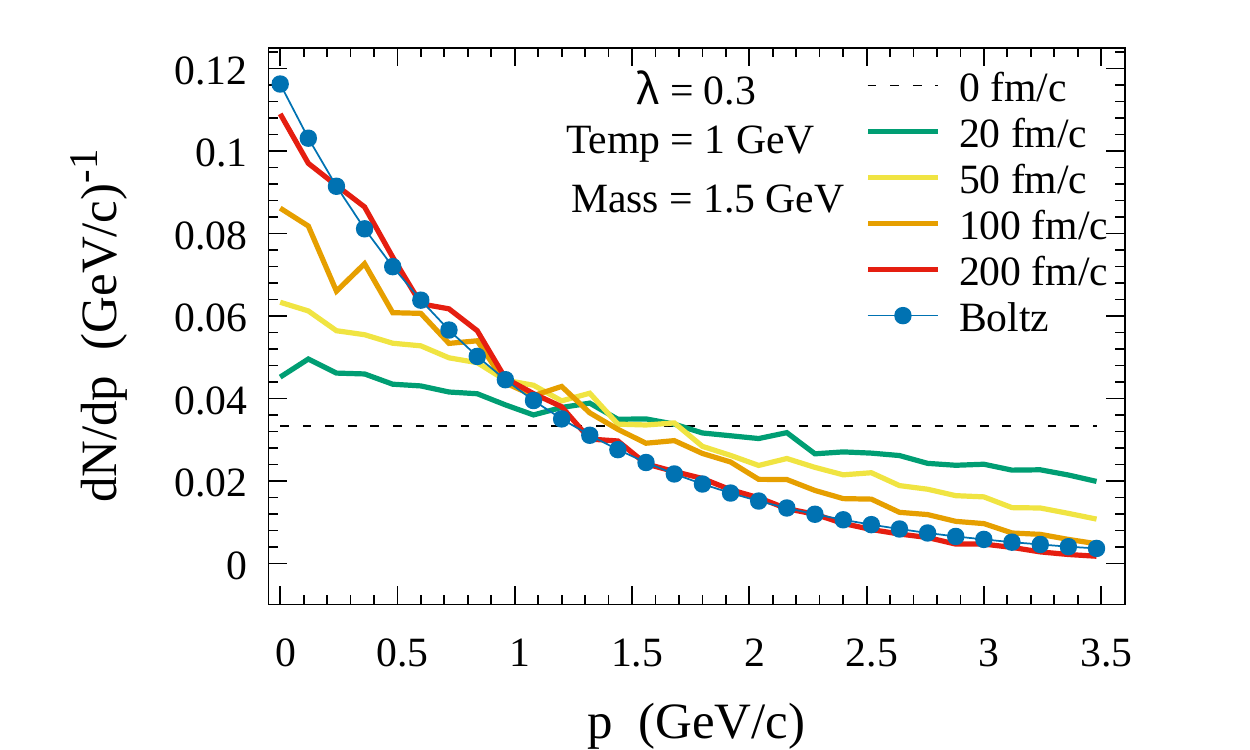}
\caption{Time evolution of heavy quark momentum distribution at different 
times. The normalized distribution $dN/dp$ 
represents the probability of heavy quarks located at the momentum $p$. 
Heavy quark mass is chosen as $m=1.5$ GeV. The temperature of the medium is $T=1$ GeV. 
The coupling strength is taken to be $\lambda=0.3$. 
Initial momentum distribution of heavy quark 
is taken to be an uniform distribution. Blue-marked line is the Boltzmann distribution 
which is the thermalization limit. 
}
\label{lab-fig-couple03}
\end{figure}

In Fig.\ref{lab-fig-couple03}, 
we initialize the wave function of heavy quarks with an uniform distribution. 
The momentum distribution of heavy quarks at different times 
are plotted with different colors. The effect of the 
damping factor $e^{-{\beta\over 2}\epsilon_{n_{ij}}}$ which comes from the part of the 
medium wave function, reduce the density of heavy quark 
at high momentum states while increase 
the density at low momentum eigenstates.  
The term with random phase factor make heavy quarks uniformly distribute over 
all the momentum eignestates. The combined effect of two terms make the system 
evolve towards the distribution of the medium.

\begin{figure}[h]
\includegraphics[width=0.49\textwidth]{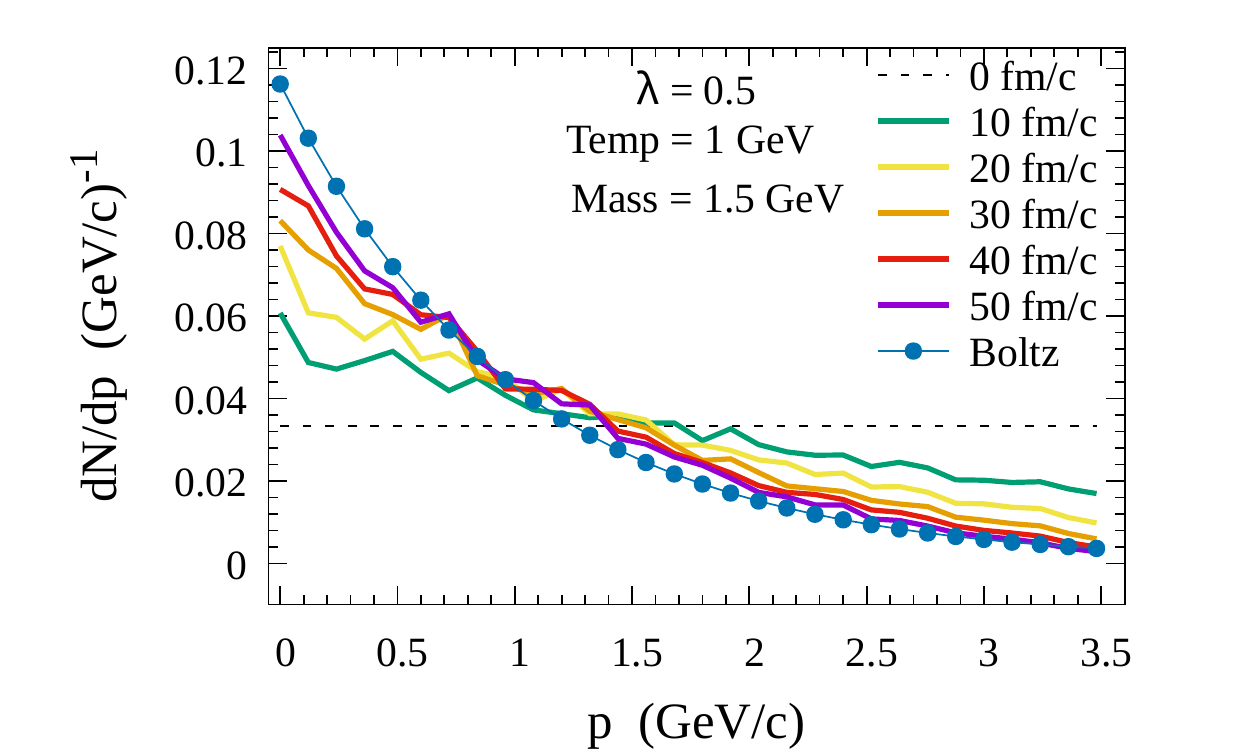}
\caption{Time evolution of heavy quark momentum distribution $dN/dp$ at different times. 
The coupling strength is taken as $\lambda=0.5$. Other parameters are the same 
with Fig.\ref{lab-fig-couple03}. 
}
\label{lab-fig-couple05}
\end{figure}

In order to study the effects of the coupling strength on the thermalization 
process, 
we take $\lambda=0.5$ in Fig.\ref{lab-fig-couple05}. 
The thermalization time with stronger coupling 
strength becomes much shorter than the case in 
Fig.\ref{lab-fig-couple03} ~\cite{Moore:2004tg,Zhao:2020jqu}. 
The thermalization process 
also depends on the density of thermal particles in the environment. 
In the case of heavy-ion collisions, 
the realistic initial momentum distribution of heavy quarks and the density of the medium 
should be considered. 
By employing a realistic momentum distribution instead 
of the uniform distribution and the , the thermalization time is also much shorter than the 
case in the figures. 
As the motivation 
of this work is not trying to compare with experimental data about heavy quarks, 
instead, we intend to 
build the SSE model to treat heavy quarks as an open quantum system. 
In the future work, we will apply 
this open quantum system approach to the realistic evolutions of heavy quarks and 
to the case of heavy quarkonium in heavy-ion collisions.  

\section{Summary}
In this work, we treat heavy quarks as an open quantum system in the medium 
and obtain the SSE for heavy quarks from full 
Schr\"odinger equation. 
The random interactions 
between heavy quark (system) and the medium (environment) are included via the stochastic 
potential, which exchange the momentum and phase factor between wave functions of heavy quarks 
and the medium. As a preliminary study, we initialize the momentum distribution of 
heavy quarks with an uniform distribution, while the static 
medium consisting of massless particles 
satisfies a Boltzmann distribution. A large set of stochastic wave functions 
of heavy quarks are obtained by solving the SSE along different quantum trajectories. 
After the ensemble average, the mean wave function and the momentum 
distribution of heavy quarks 
at different times are presented, which evolves toward the thermal 
distribution of the 
medium. 
Thermalization process with different coupling strength between the system and the 
environment are also studies. 
In the future work, we will apply this SSE model to the realistic 
evolutions of heavy quarks and extend the model to the quarkonium case 
in heavy-ion collisions.

\vspace{0.5cm}
{\bf Acknowledgement:}
This work is supported by the
National Natural Science Foundation of China (NSFC)
under Grant Nos. 12175165, 11705125.


\end{spacing}

\begin{thebibliography}{20}

\bibitem{Bazavov:2011nk}
A.~Bazavov, \textit{et al.}
Phys. Rev. D \textbf{85}, 054503 (2012)
doi:10.1103/PhysRevD.85.054503
[arXiv:1111.1710 [hep-lat]].

\bibitem{Matsui:1986dk}
T.~Matsui and H.~Satz,
Phys. Lett. B \textbf{178}, 416-422 (1986)
doi:10.1016/0370-2693(86)91404-8

\bibitem{Thews:2000rj}
R.~L.~Thews, M.~Schroedter and J.~Rafelski,
Phys. Rev. C \textbf{63}, 054905 (2001)
doi:10.1103/PhysRevC.63.054905
[arXiv:hep-ph/0007323 [hep-ph]].

\bibitem{Greco:2003vf}
V.~Greco, C.~M.~Ko and R.~Rapp,
Phys. Lett. B \textbf{595}, 202-208 (2004)
doi:10.1016/j.physletb.2004.06.064
[arXiv:nucl-th/0312100 [nucl-th]].


\bibitem{Grandchamp:2003uw}
L.~Grandchamp, R.~Rapp and G.~E.~Brown,
Phys. Rev. Lett. \textbf{92}, 212301 (2004)
doi:10.1103/PhysRevLett.92.212301
[arXiv:hep-ph/0306077 [hep-ph]].


\bibitem{Yan:2006ve}
L.~Yan, P.~Zhuang and N.~Xu,
Phys. Rev. Lett. \textbf{97}, 232301 (2006)
doi:10.1103/PhysRevLett.97.232301
[arXiv:nucl-th/0608010 [nucl-th]].


\bibitem{Liu:2010ej}
Y.~Liu, B.~Chen, N.~Xu and P.~Zhuang,
Phys. Lett. B \textbf{697}, 32-36 (2011)
doi:10.1016/j.physletb.2011.01.026
[arXiv:1009.2585 [nucl-th]].


\bibitem{Blaizot:2017ypk}
J.~P.~Blaizot and M.~A.~Escobedo,
JHEP \textbf{06}, 034 (2018)
doi:10.1007/JHEP06(2018)034
[arXiv:1711.10812 [hep-ph]].


\bibitem{Chen:2019qzx}
B.~Chen, M.~Hu, H.~Zhang and J.~Zhao,
Phys. Lett. B \textbf{802}, 135271 (2020)
doi:10.1016/j.physletb.2020.135271
[arXiv:1910.08275 [nucl-th]].

\bibitem{Yao:2018sgn}
X.~Yao and B.~M\"uller,
Phys. Rev. D \textbf{100}, no.1, 014008 (2019)
doi:10.1103/PhysRevD.100.014008
[arXiv:1811.09644 [hep-ph]].
\bibitem{Yao:2020xzw}
X.~Yao, W.~Ke, Y.~Xu, S.~A.~Bass and B.~M\"uller,
JHEP \textbf{01}, 046 (2021)
doi:10.1007/JHEP01(2021)046
[arXiv:2004.06746 [hep-ph]].

\bibitem{vanHees:2007me}
H.~van Hees, M.~Mannarelli, V.~Greco and R.~Rapp,
Phys. Rev. Lett. \textbf{100}, 192301 (2008)
doi:10.1103/PhysRevLett.100.192301
[arXiv:0709.2884 [hep-ph]].


\bibitem{Zhao:2017yan}
J.~Zhao and B.~Chen,
Phys. Lett. B \textbf{776}, 17-21 (2018)
doi:10.1016/j.physletb.2017.11.014
[arXiv:1705.04558 [nucl-th]].


\bibitem{Rapp:2018qla}
R.~Rapp, P.~B.~Gossiaux, A.~Andronic, R.~Averbeck, S.~Masciocchi, A.~Beraudo, E.~Bratkovskaya, P.~Braun-Munzinger, S.~Cao and A.~Dainese, \textit{et al.}
Nucl. Phys. A \textbf{979}, 21-86 (2018)
doi:10.1016/j.nuclphysa.2018.09.002
[arXiv:1803.03824 [nucl-th]].

\bibitem{Qin:2010mn}
G.~Y.~Qin and B.~Muller,
Phys. Rev. Lett. \textbf{106}, 162302 (2011)
[erratum: Phys. Rev. Lett. \textbf{108}, 189904 (2012)]
doi:10.1103/PhysRevLett.106.162302
[arXiv:1012.5280 [hep-ph]].

\bibitem{Baier:1996kr}
R.~Baier, Y.~L.~Dokshitzer, A.~H.~Mueller, S.~Peigne and D.~Schiff,
Nucl. Phys. B \textbf{483}, 291-320 (1997)
doi:10.1016/S0550-3213(96)00553-6
[arXiv:hep-ph/9607355 [hep-ph]].

\bibitem{Qin:2007rn}
G.~Y.~Qin, J.~Ruppert, C.~Gale, S.~Jeon, G.~D.~Moore and M.~G.~Mustafa,
Phys. Rev. Lett. \textbf{100}, 072301 (2008)
doi:10.1103/PhysRevLett.100.072301
[arXiv:0710.0605 [hep-ph]].

\bibitem{He:2011qa}
M.~He, R.~J.~Fries and R.~Rapp,
Phys. Rev. C \textbf{86}, 014903 (2012)
doi:10.1103/PhysRevC.86.014903

\bibitem{Cao:2013ita}
S.~Cao, G.~Y.~Qin and S.~A.~Bass,
Phys. Rev. C \textbf{88}, 044907 (2013)
doi:10.1103/PhysRevC.88.044907
[arXiv:1308.0617 [nucl-th]].


\bibitem{Cao:2016gvr}
S.~Cao, T.~Luo, G.~Y.~Qin and X.~N.~Wang,
Phys. Rev. C \textbf{94}, no.1, 014909 (2016)
doi:10.1103/PhysRevC.94.014909
[arXiv:1605.06447 [nucl-th]].

\bibitem{ref:qsm1}
Bassano Vacchini, Klaus Hornberger, 
Physics Reports 478 (2009) 71-120 [arXiv:0904.3911[quant-ph]]. 


\bibitem{Brambilla:2020qwo}
N.~Brambilla, M.~\'A.~Escobedo, M.~Strickland, A.~Vairo, P.~Vander Griend and J.~H.~Weber,
JHEP \textbf{05}, 136 (2021)
doi:10.1007/JHEP05(2021)136
[arXiv:2012.01240 [hep-ph]].

\bibitem{Katz:2015qja}
R.~Katz and P.~B.~Gossiaux,
Annals Phys. \textbf{368}, 267-295 (2016)
doi:10.1016/j.aop.2016.02.005
[arXiv:1504.08087 [quant-ph]].

\bibitem{Akamatsu:2018xim}
Y.~Akamatsu, M.~Asakawa, S.~Kajimoto and A.~Rothkopf,
JHEP \textbf{07}, 029 (2018)
doi:10.1007/JHEP07(2018)029
[arXiv:1805.00167 [nucl-th]].

\bibitem{He:2012df}
M.~He, R.~J.~Fries and R.~Rapp,
Phys. Rev. Lett. \textbf{110}, no.11, 112301 (2013)
doi:10.1103/PhysRevLett.110.112301
[arXiv:1204.4442 [nucl-th]].


\bibitem{Chen:2017duy}
B.~Chen and J.~Zhao,
Phys. Lett. B \textbf{772}, 819-824 (2017)
doi:10.1016/j.physletb.2017.07.054
[arXiv:1704.05622 [nucl-th]].

\bibitem{ref:sse1}
R Biele and R D’Agosta, J. Phys.: Condens. Matter 24 (2012) 273201.
doi:10.1088/0953-8984/24/27/273201.


\bibitem{ref:project1}
Robert Zwanzig,  J. Chem. Phys. 33, 1338 (1960). 
doi: 10.1063/1.1731409

\bibitem{ref:project2}
P. Gaspard and M. Nagaoka, J. Chem. Phys. 111, 5676 (1999). doi: 10.1063/1.479868

\bibitem{Kaczmarek:2002mc}
O.~Kaczmarek, F.~Karsch, P.~Petreczky and F.~Zantow,
Phys. Lett. B \textbf{543}, 41-47 (2002)
doi:10.1016/S0370-2693(02)02415-2
[arXiv:hep-lat/0207002 [hep-lat]].

\bibitem{Lafferty:2019jpr}
D.~Lafferty and A.~Rothkopf,
Phys. Rev. D \textbf{101}, no.5, 056010 (2020)
doi:10.1103/PhysRevD.101.056010
[arXiv:1906.00035 [hep-ph]].

\bibitem{Moore:2004tg}
G.~D.~Moore and D.~Teaney,
Phys. Rev. C \textbf{71}, 064904 (2005)
doi:10.1103/PhysRevC.71.064904
[arXiv:hep-ph/0412346 [hep-ph]].

\bibitem{Zhao:2020jqu}
J.~Zhao, K.~Zhou, S.~Chen and P.~Zhuang,
Prog. Part. Nucl. Phys. \textbf{114}, 103801 (2020)
doi:10.1016/j.ppnp.2020.103801
[arXiv:2005.08277 [nucl-th]].

%



\end{thebibliography}
\end{document}